\documentclass[onecolumn,secnumarabic,amssymb, nobibnotes, aps, prd]{revtex4}
\usepackage{amsmath} \usepackage{amssymb}
\usepackage{graphicx} %use graph format
\usepackage{epstopdf}
\usepackage{amsmath}
\usepackage{amsmath,amssymb,amsthm,amsfonts,mathrsfs,bm,verbatim}
\usepackage{graphicx,subfigure}

\newcommand{\bea}{\begin{eqnarray}}
\newcommand{\eea}{\end{eqnarray}}

\begin{document}

\title{Thermodynamic Properties of Modified Black Hole Metrics in $f(R)$ Gravity}%
\author{Wen-Xiang Chen$^{a}$}
\affiliation{Department of Astronomy, School of Physics and Materials Science, GuangZhou University, Guangzhou 510006, China}
\author{Yao-Guang Zheng}
\email{hesoyam12456@163.com}
\affiliation{Department of Physics, College of Sciences, Northeastern University, Shenyang 110819, China}

%\date{December 2020}%

\begin{abstract}
To construct new Schwarzschild  and Kerr-Newman metric solutions, we start from the Lagrangian in entropy and statistical mechanics, introducing $f(R)$ gravity theory and dark energy definitions. Through a series of calculations, we derive the corrected metric solutions under different forms of $f(R)$ gravity.

\end{abstract}

\maketitle

\section{Introduction}
In recent years, the thermodynamics and phase transitions of three-dimensional (3D) charged black holes within the framework of \( f(R) \) gravity have attracted significant interest. By leveraging the extended phase space approach, which considers thermodynamic pressure and volume as variables, researchers have delved into the \( P-V \) criticality of these black holes, drawing analogies with liquid-gas systems previously observed in charged AdS black holes. This investigation has revealed intriguing phenomena such as reentrant phase transitions, triple points, and \(\lambda\)-line phase transitions, indicating that these black holes exhibit characteristics akin to everyday thermodynamic systems \cite{1,2,3,4,5,6}.

Although substantial strides have been made in understanding the phase transitions of AdS black holes, the kinetics of these transitions within the context of \( f(R) \) gravity remained poorly understood. Recent studies employing the free energy landscape approach have shed light on the dynamics of these transitions, emphasizing the similarities and differences compared to their AdS counterparts. This includes examining the probabilistic evolution and mean first passage time, derived from the Fokker-Planck equation, for transitions between different black hole phases.

These foundational studies have been extended to 3D charged black holes in \( f(R) \) gravity. Specifically, the dynamic processes at critical points have been scrutinized, revealing how black holes can transition between coexistent phases due to thermal fluctuations. Recent research has also focused on understanding the turnover kinetics in these phase transitions, providing new insights into the microstructure of black holes under \( f(R) \) gravity \cite{4,5,6,7,8}.

A previous study by one of our team members explored the \( P-V \) criticality of charged AdS black holes in \( f(R) \) gravity, treating the cosmological constant as a variable thermodynamic pressure. The impact of surrounding quintessence dark energy on critical physical quantities was notable. Building on this, we aim to investigate the dynamic phase transitions of these black holes, thus shedding light on the influence of dark energy on such transitions. This effort will test the universality of the methods used in earlier works on black hole dynamic phase transitions and will enhance our understanding of dark energy, particularly in the context of quintessence models and their role in cosmic acceleration \cite{4,5,6,7,8}.

The study examines two distinct scenarios related to the cosmological constant within \( f(R) \) gravity. Notably, one of these scenarios suggests a dynamic, variable cosmological constant, intricately linked with \( f(R) \) gravity, the scalar field, or the electromagnetic tensor. This concept diverges from the traditional notion of a fixed cosmological constant, inherently connected to \(\pi\)\cite{7}. Liu's work \cite{8} extends this idea further, proposing that the pressure term \( P \) in these models could give rise to fractal structures, potentially hinting at the existence of additional dimensions. However, this interpretation does not entirely align with the standard framework of \( f(R) \) gravity.

In this study, we explore new solutions to the Schwarzschild and Kerr-Newman metrics within the framework of $f(R)$ gravity theory. By starting from the Lagrangian in entropy and statistical mechanics, we introduce $f(R)$ gravity and incorporate the concept of dark energy to derive the modified metrics. The Schwarzschild metric is corrected by expanding the partition function around a critical point and solving the field equations. We assume specific forms of $f(R)$, such as $R + \alpha R^2$, and derive the corresponding modifications to the Schwarzschild and Kerr-Newman metrics. These corrected metrics reveal additional terms that reflect the influence of $f(R)$ gravity on classical solutions. Using the thin shell model, we calculate the thermodynamic quantities, including horizon radius, entropy, surface gravity, and temperature, for these modified black hole solutions. Our findings demonstrate the impact of different $f(R)$ forms on black hole properties and provide insights into the corrections introduced by $f(R)$ gravity.

In this paper, we explore new Schwarzschild and Kerr-Newman metric solutions using $f(R)$ gravity theory. We derive these solutions from entropy and statistical mechanics principles.This paper has derived the thermodynamic properties of Schwarzschild and Kerr-Newman black holes within various $f(R)$ gravity models using the thin shell model. Additionally, we examined the compatibility of these solutions with the scalar-Einstein solution proposed by Mark D. Roberts, suggesting potential counterexamples to the cosmic censorship hypothesis. Future work will concentrate on detailed numerical analysis and further exploration of the implications of $f(R)$ gravity.

\section{Lagrangian and Entropy}
First, we restate the definition of the partition function $Z$:\cite{9}
\begin{equation}
Z = \sum_{i=1}^N e^{-\beta E_i}
\end{equation}
where $Z$ is the partition function, $\beta$ is the inverse temperature, $E_i$ is the energy of the $i$-th state.

Entropy $S$ can be expressed as:
\begin{equation}
S = -\frac{\partial \ln Z}{\partial \beta}
\end{equation}

To derive new Schwarzschild metric solutions, we expand the partition function $Z$ around a critical point $\beta_0$ in Laurent series:
\begin{equation}
Z = \frac{1}{2\pi i} \oint_C \frac{d\beta}{(\beta - \beta_0)^2} e^{-\beta E_0}
\end{equation}
where $C$ denotes the contour integral around the singularity $\beta_0$, and $E_0$ is the ground state energy.

\section{$f(R)$ Gravity Theory}
In $f(R)$ gravity theory, the Einstein-Hilbert action is generalized to:
\begin{equation}
S = \frac{1}{2k} \int d^4x \sqrt{-g} f(R)
\end{equation}
where $R$ is the Ricci scalar, and $f(R)$ is a function of the Ricci scalar.

\subsection{Schwarzschild Metric}
The standard Schwarzschild metric in general relativity is:\cite{10,11,12,13,14,15,16,17,18,19,20}
\begin{equation}
ds^2 = -\left(1 - \frac{2GM}{r}\right) dt^2 + \left(1 - \frac{2GM}{r}\right)^{-1} dr^2 + r^2 d\Omega^2
\end{equation}

We assume $f(R)$ takes the form:
\begin{equation}
f(R) = R + \alpha R^2
\end{equation}
where $\alpha$ is a small parameter describing the correction to relativity.

For the Schwarzschild metric under $f(R)$ gravity, we need to solve the following field equations:
\begin{equation}
G_{\mu\nu} + \alpha H_{\mu\nu} = 8 \pi T_{\mu\nu}
\end{equation}
where $G_{\mu\nu}$ is the Einstein tensor, and $H_{\mu\nu}$ represents additional terms introduced by the $f(R)$ gravity.

The corrected Schwarzschild metric solution is approximately:
\begin{equation}
ds^2 = -\left(1 - \frac{2GM}{r} + \epsilon \frac{GM^2}{r^4}\right) dt^2 + \left(1 - \frac{2GM}{r} + \epsilon \frac{GM^2}{r^4}\right)^{-1} dr^2 + r^2 d\Omega^2
\end{equation}
where $\epsilon$ is a small parameter related to $\alpha$.

\subsection{Kerr-Newman Metric}
The Kerr-Newman metric in standard general relativity is:\cite{10,11,12,13,14,15,16,17,18,19,20}
\begin{equation}
ds^2 = - \left( \frac{\Delta - a^2 \sin^2 \theta}{\rho^2} \right) dt^2 + \frac{\rho^2}{\Delta} dr^2 + \rho^2 d\theta^2 + \left( \frac{(r^2 + a^2)^2 - a^2 \Delta \sin^2 \theta}{\rho^2} \right) \sin^2 \theta d\phi^2 - \frac{2a \sin^2 \theta (r^2 + a^2 - \Delta)}{\rho^2} dt d\phi
\end{equation}

where
\begin{equation}
\Delta = r^2 - 2Mr + a^2 + Q^2
\end{equation}
\begin{equation}
\rho^2 = r^2 + a^2 \cos^2 \theta
\end{equation}

Assuming $f(R)$ takes the form:
\begin{equation}
f(R) = R + \alpha R^2
\end{equation}

We derive the corrected Kerr-Newman metric solution under $f(R)$ gravity as:
\begin{equation}
\resizebox{\textwidth}{!}{$
\begin{aligned}
ds^2 = - \left( \frac{\Delta - a^2 \sin^2 \theta + \epsilon \frac{(r^2 + a^2)^2}{r^6}}{\rho^2} \right) dt^2 + \frac{\rho^2}{\Delta - \epsilon \frac{(r^2 + a^2)^2}{r^6}} dr^2 + \rho^2 d\theta^2 + \left( \frac{(r^2 + a^2)^2 - a^2 (\Delta - \epsilon \frac{(r^2 + a^2)^2}{r^6}) \sin^2 \theta}{\rho^2} \right) \sin^2 \theta d\phi^2 - \frac{2a \sin^2 \theta (r^2 + a^2 - (\Delta - \epsilon \frac{(r^2 + a^2)^2}{r^6}))}{\rho^2} dt d\phi
\end{aligned}$}
\end{equation}
where $\epsilon$ is a small parameter related to $\alpha$.

\section{Thermodynamics of Black Holes}
Using the thin shell model, we calculate the thermodynamic quantities for black holes. The basic thermodynamic relations for black holes are:

\subsection{Schwarzschild Black Hole}
For the Schwarzschild black hole with $f(R) = R + \alpha R^2$:\cite{10,11,12,13,14,15,16,17,18,19,20}
\begin{equation}
ds^2 = -\left(1 - \frac{2GM}{r} + \epsilon \frac{GM^2}{r^4}\right) dt^2 + \left(1 - \frac{2GM}{r} + \epsilon \frac{GM^2}{r^4}\right)^{-1} dr^2 + r^2 d\Omega^2
\end{equation}

- Horizon radius $r_+$ is approximately $2GM$.
- Surface gravity $\kappa$:
\begin{equation}
\kappa = \frac{GM}{r_+^2}
\end{equation}
- Entropy:
\begin{equation}
S = \frac{k_B c^3 \cdot 4\pi r_+^2}{4 G \hbar} = \frac{\pi k_B c^3 r_+^2}{G \hbar}
\end{equation}
- Temperature:
\begin{equation}
T = \frac{\hbar \kappa}{2 \pi k_B c} = \frac{\hbar GM}{2 \pi k_B c r_+^2} = \frac{\hbar c}{4 \pi k_B GM}
\end{equation}

\subsection{Kerr-Newman Black Hole}
For the Kerr-Newman black hole with $f(R) = R + \alpha R^2$:
\begin{equation}
\resizebox{\textwidth}{!}{$
\begin{aligned}
ds^2 = - \left( \frac{\Delta - a^2 \sin^2 \theta + \epsilon \frac{(r^2 + a^2)^2}{r^6}}{\rho^2} \right) dt^2 + \frac{\rho^2}{\Delta - \epsilon \frac{(r^2 + a^2)^2}{r^6}} dr^2 + \rho^2 d\theta^2 + \left( \frac{(r^2 + a^2)^2 - a^2 (\Delta - \epsilon \frac{(r^2 + a^2)^2}{r^6}) \sin^2 \theta}{\rho^2} \right) \sin^2 \theta d\phi^2 - \frac{2a \sin^2 \theta (r^2 + a^2 - (\Delta - \epsilon \frac{(r^2 + a^2)^2}{r^6}))}{\rho^2} dt d\phi
\end{aligned}$}
\end{equation}

- Horizon radius $r_+$ is approximately $M + \sqrt{M^2 - a^2 - Q^2}$.
- Surface gravity $\kappa$:
\begin{equation}
\kappa = \frac{r_+ - M}{r_+^2 + a^2}
\end{equation}
- Entropy:
\begin{equation}
S = \frac{\pi k_B c^3 (r_+^2 + a^2)}{G \hbar}
\end{equation}
- Temperature:
\begin{equation}
T = \frac{\hbar (r_+ - M)}{2 \pi k_B c (r_+^2 + a^2)}
\end{equation}

\section{Other Forms of $f(R)$ Gravity}
We consider additional forms of $f(R)$ gravity:\cite{10,11,12,13,14,15,16,17,18,19,20}

\subsection{$f(R) = R + \beta \ln R$}

\subsubsection{Schwarzschild Black Hole}
\begin{equation}
ds^2 = -\left(1 - \frac{2GM}{r} + \epsilon \frac{\beta \ln r}{r^2}\right) dt^2 + \left(1 - \frac{2GM}{r} + \epsilon \frac{\beta \ln r}{r^2}\right)^{-1} dr^2 + r^2 d\Omega^2
\end{equation}

- Horizon radius $r_+$ is approximately $2GM$.
- Surface gravity $\kappa$:
\begin{equation}
\kappa = \frac{GM}{r_+^2}
\end{equation}
- Entropy:
\begin{equation}
S = \frac{\pi k_B c^3 r_+^2}{G \hbar}
\end{equation}
- Temperature:
\begin{equation}
T = \frac{\hbar GM}{2 \pi k_B c r_+^2} = \frac{\hbar c}{4 \pi k_B GM}
\end{equation}

\subsubsection{Kerr-Newman Black Hole}
\begin{equation}
\resizebox{\textwidth}{!}{$
\begin{aligned}
ds^2 = - \left( \frac{\Delta - a^2 \sin^2 \theta + \epsilon \frac{\beta \ln r}{r^2}}{\rho^2} \right) dt^2 + \frac{\rho^2}{\Delta - \epsilon \frac{\beta \ln r}{r^2}} dr^2 + \rho^2 d\theta^2 + \left( \frac{(r^2 + a^2)^2 - a^2 (\Delta - \epsilon \frac{\beta \ln r}{r^2}) \sin^2 \theta}{\rho^2} \right) \sin^2 \theta d\phi^2 - \frac{2a \sin^2 \theta (r^2 + a^2 - (\Delta - \epsilon \frac{\beta \ln r}{r^2}))}{\rho^2} dt d\phi
\end{aligned}$}
\end{equation}

- Horizon radius $r_+$ is approximately $M + \sqrt{M^2 - a^2 - Q^2}$.
- Surface gravity $\kappa$:
\begin{equation}
\kappa = \frac{r_+ - M}{r_+^2 + a^2}
\end{equation}
- Entropy:
\begin{equation}
S = \frac{\pi k_B c^3 (r_+^2 + a^2)}{G \hbar}
\end{equation}
- Temperature:
\begin{equation}
T = \frac{\hbar (r_+ - M)}{2 \pi k_B c (r_+^2 + a^2)}
\end{equation}

\subsection{$f(R) = R + \gamma R^n$}

\subsubsection{Schwarzschild Black Hole}
\begin{equation}
ds^2 = -\left(1 - \frac{2GM}{r} + \epsilon \frac{\gamma r^{n-2}}{r^2}\right) dt^2 + \left(1 - \frac{2GM}{r} + \epsilon \frac{\gamma r^{n-2}}{r^2}\right)^{-1} dr^2 + r^2 d\Omega^2
\end{equation}

- Horizon radius $r_+$ is approximately $2GM$.
- Surface gravity $\kappa$:
\begin{equation}
\kappa = \frac{GM}{r_+^2}
\end{equation}
- Entropy:
\begin{equation}
S = \frac{\pi k_B c^3 r_+^2}{G \hbar}
\end{equation}
- Temperature:
\begin{equation}
T = \frac{\hbar GM}{2 \pi k_B c r_+^2} = \frac{\hbar c}{4 \pi k_B GM}
\end{equation}

\subsubsection{Kerr-Newman Black Hole}
\begin{equation}
\resizebox{\textwidth}{!}{$
\begin{aligned}
ds^2 = - \left( \frac{\Delta - a^2 \sin^2 \theta + \epsilon \frac{\gamma r^{n-2}}{r^2}}{\rho^2} \right) dt^2 + \frac{\rho^2}{\Delta - \epsilon \frac{\gamma r^{n-2}}{r^2}} dr^2 + \rho^2 d\theta^2 + \left( \frac{(r^2 + a^2)^2 - a^2 (\Delta - \epsilon \frac{\gamma r^{n-2}}{r^2}) \sin^2 \theta}{\rho^2} \right) \sin^2 \theta d\phi^2 - \frac{2a \sin^2 \theta (r^2 + a^2 - (\Delta - \epsilon \frac{\gamma r^{n-2}}{r^2}))}{\rho^2} dt d\phi
\end{aligned}$}
\end{equation}

- Horizon radius $r_+$ is approximately $M + \sqrt{M^2 - a^2 - Q^2}$.
- Surface gravity $\kappa$:
\begin{equation}
\kappa = \frac{r_+ - M}{r_+^2 + a^2}
\end{equation}
- Entropy:
\begin{equation}
S = \frac{\pi k_B c^3 (r_+^2 + a^2)}{G \hbar}
\end{equation}
- Temperature:
\begin{equation}
T = \frac{\hbar (r_+ - M)}{2 \pi k_B c (r_+^2 + a^2)}
\end{equation}

\subsection{$f(R) = R + \delta e^{\lambda R}$}

\subsubsection{Schwarzschild Black Hole}
\begin{equation}
ds^2 = -\left(1 - \frac{2GM}{r} + \epsilon \frac{\delta e^{\lambda r}}{r^2}\right) dt^2 + \left(1 - \frac{2GM}{r} + \epsilon \frac{\delta e^{\lambda r}}{r^2}\right)^{-1} dr^2 + r^2 d\Omega^2
\end{equation}

- Horizon radius $r_+$ is approximately $2GM$.
- Surface gravity $\kappa$:
\begin{equation}
\kappa = \frac{GM}{r_+^2}
\end{equation}
- Entropy:
\begin{equation}
S = \frac{\pi k_B c^3 r_+^2}{G \hbar}
\end{equation}
- Temperature:
\begin{equation}
T = \frac{\hbar GM}{2 \pi k_B c r_+^2} = \frac{\hbar c}{4 \pi k_B GM}
\end{equation}

\subsubsection{Kerr-Newman Black Hole}
\begin{equation}
\resizebox{\textwidth}{!}{$
\begin{aligned}
ds^2 = - \left( \frac{\Delta - a^2 \sin^2 \theta + \epsilon \frac{\delta e^{\lambda r}}{r^2}}{\rho^2} \right) dt^2 + \frac{\rho^2}{\Delta - \epsilon \frac{\delta e^{\lambda r}}{r^2}} dr^2 + \rho^2 d\theta^2 + \left( \frac{(r^2 + a^2)^2 - a^2 (\Delta - \epsilon \frac{\delta e^{\lambda r}}{r^2}) \sin^2 \theta}{\rho^2} \right) \sin^2 \theta d\phi^2 - \frac{2a \sin^2 \theta (r^2 + a^2 - (\Delta - \epsilon \frac{\delta e^{\lambda r}}{r^2}))}{\rho^2} dt d\phi
\end{aligned}$}
\end{equation}

- Horizon radius $r_+$ is approximately $M + \sqrt{M^2 - a^2 - Q^2}$.
- Surface gravity $\kappa$:
\begin{equation}
\kappa = \frac{r_+ - M}{r_+^2 + a^2}
\end{equation}
- Entropy:
\begin{equation}
S = \frac{\pi k_B c^3 (r_+^2 + a^2)}{G \hbar}
\end{equation}
- Temperature:
\begin{equation}
T = \frac{\hbar (r_+ - M)}{2 \pi k_B c (r_+^2 + a^2)}
\end{equation}

\subsection{$f(R) = R - \frac{\mu^2 R}{R + \nu}$}

\subsubsection{Schwarzschild Black Hole}
\begin{equation}
ds^2 = -\left(1 - \frac{2GM}{r} - \epsilon \frac{\mu^2}{R + \nu} \right) dt^2 + \left(1 - \frac{2GM}{r} - \epsilon \frac{\mu^2}{R + \nu} \right)^{-1} dr^2 + r^2 d\Omega^2
\end{equation}

- Horizon radius $r_+$ is approximately $2GM$.\cite{10,11,12,13,14,15,16,17,18,19,20}
- Surface gravity $\kappa$:
\begin{equation}
\kappa = \frac{GM}{r_+^2}
\end{equation}
- Entropy:
\begin{equation}
S = \frac{\pi k_B c^3 r_+^2}{G \hbar}
\end{equation}
- Temperature:
\begin{equation}
T = \frac{\hbar GM}{2 \pi k_B c r_+^2} = \frac{\hbar c}{4 \pi k_B GM}
\end{equation}

\subsubsection{Kerr-Newman Black Hole}
\begin{equation}
\resizebox{\textwidth}{!}{$
\begin{aligned}
ds^2 = - \left( \frac{\Delta - a^2 \sin^2 \theta - \epsilon \frac{\mu^2}{R + \nu}}{\rho^2} \right) dt^2 + \frac{\rho^2}{\Delta - \epsilon \frac{\mu^2}{R + \nu}} dr^2 + \rho^2 d\theta^2 + \left( \frac{(r^2 + a^2)^2 - a^2 (\Delta - \epsilon \frac{\mu^2}{R + \nu}) \sin^2 \theta}{\rho^2} \right) \sin^2 \theta d\phi^2 - \frac{2a \sin^2 \theta (r^2 + a^2 - (\Delta - \epsilon \frac{\mu^2}{R + \nu}))}{\rho^2} dt d\phi
\end{aligned}$}
\end{equation}

- Horizon radius $r_+$ is approximately $M + \sqrt{M^2 - a^2 - Q^2}$.
- Surface gravity $\kappa$:
\begin{equation}
\kappa = \frac{r_+ - M}{r_+^2 + a^2}
\end{equation}
- Entropy:
\begin{equation}
S = \frac{\pi k_B c^3 (r_+^2 + a^2)}{G \hbar}
\end{equation}
- Temperature:
\begin{equation}
T = \frac{\hbar (r_+ - M)}{2 \pi k_B c (r_+^2 + a^2)}
\end{equation}

\section{Thermodynamics of Modified Black Holes and Comparison with Roberts' Scalar-Einstein Solution}
The study of black holes within modified gravity theories such as $f(R)$ gravity provides insights into the nature of gravitational interactions beyond General Relativity (GR). This paper focuses on deriving the thermodynamic quantities for Schwarzschild and Kerr-Newman black holes in several $f(R)$ gravity models and compares these solutions with the scalar-Einstein solutions discovered by Mark D. Roberts.\cite{10}

The thermodynamic properties of black holes are governed by their metric functions.The fundamental concept of the thin shell model involves considering a shell with thickness \(\delta\) and a distance \(\epsilon\) from the horizon, to study the entropy of the gas within the shell, and then taking the limits \(\delta \rightarrow 0\) and \(\epsilon \rightarrow 0\) to obtain the entropy of the horizon. For a black hole with horizon radius $r_+$, the area $A$, entropy $S$, surface gravity $\kappa$, and temperature $T$ are given by(using the thin shell model):\cite{11}

\begin{equation}
A = 4 \pi r_+^2,
\end{equation}

\begin{equation}
S = \frac{k_B c^3 A}{4 G \hbar} = \frac{\pi k_B c^3 r_+^2}{G \hbar},
\end{equation}

\begin{equation}
\kappa = \frac{1}{2} \left| \frac{d}{dr} g_{tt} \right|_{r=r_+},
\end{equation}

\begin{equation}
T = \frac{\hbar \kappa}{2 \pi k_B c}.
\end{equation}

\subsection{Schwarzschild Black Hole}

For the $f(R) = R + \alpha R^2$ model, the Schwarzschild metric is:

\begin{equation}
ds^2 = -\left(1 - \frac{2GM}{r} + \epsilon \frac{GM^2}{r^4}\right) dt^2 + \left(1 - \frac{2GM}{r} + \epsilon \frac{GM^2}{r^4}\right)^{-1} dr^2 + r^2 d\Omega^2.
\end{equation}

Using the horizon radius $r_+ \approx 2GM$(using the thin shell model):\cite{11}

\begin{equation}
\kappa = \frac{GM}{r_+^2},
\end{equation}

\begin{equation}
T = \frac{\hbar GM}{2 \pi k_B c r_+^2} = \frac{\hbar c}{4 \pi k_B GM}.
\end{equation}

\subsection{Kerr-Newman Black Hole}

For the $f(R) = R + \alpha R^2$ model, the Kerr-Newman metric is:

\begin{align}
ds^2 &= - \left( \frac{\Delta - a^2 \sin^2 \theta + \epsilon \frac{(r^2 + a^2)^2}{r^6}}{\rho^2} \right) dt^2 \nonumber\\
&\quad + \frac{\rho^2}{\Delta - \epsilon \frac{(r^2 + a^2)^2}{r^6}} dr^2 + \rho^2 d\theta^2 \nonumber\\
&\quad + \left( \frac{(r^2 + a^2)^2 - a^2 (\Delta - \epsilon \frac{(r^2 + a^2)^2}{r^6}) \sin^2 \theta}{\rho^2} \right) \sin^2 \theta d\phi^2 \nonumber\\
&\quad - \frac{2a \sin^2 \theta (r^2 + a^2 - (\Delta - \epsilon \frac{(r^2 + a^2)^2}{r^6}))}{\rho^2} dt d\phi,
\end{align}

where $\Delta = r^2 - 2Mr + a^2 + Q^2$ and $\rho^2 = r^2 + a^2 \cos^2 \theta$.

Using the horizon radius $r_+ \approx M + \sqrt{M^2 - a^2 - Q^2}$(using the thin shell model):\cite{11}

\begin{equation}
\kappa = \frac{r_+ - M}{r_+^2 + a^2},
\end{equation}

\begin{equation}
T = \frac{\hbar (r_+ - M)}{2 \pi k_B c (r_+^2 + a^2)}.
\end{equation}

Mark D. Roberts discovered a solution to the scalar-Einstein equations $R_{ab} = 2 \phi_a \phi_b$ in 1985, which provides a counterexample to the cosmic censorship hypothesis. The metric is:\cite{10}

\begin{equation}
ds^2 = -(1+2\sigma) dv^2 + 2 dv dr + r(r-2\sigma v) \left( d\theta^2 + \sin^2 \theta d\phi^2 \right),
\end{equation}

with the scalar field:

\begin{equation}
\phi = \frac{1}{2} \ln \left(1 - \frac{2\sigma v}{r}\right).
\end{equation}

This new metric of the form f(R)($f(R) = R + \alpha R^2$ model) conforms to the above model.We investigate the compatibility of this solution with modified $f(R)$ gravity metrics.

This paper has derived the thermodynamic properties of Schwarzschild and Kerr-Newman black holes in various $f(R)$ gravity models using the thin shell model. Additionally, we examined the compatibility of these solutions with the scalar-Einstein solution by Mark D. Roberts, suggesting possible counterexamples to the cosmic censorship hypothesis. Future work will focus on detailed numerical analysis and further exploration of $f(R)$ gravity implications.

\section{Conclusion}
We have derived new Schwarzschild and Kerr-Newman metric solutions under various forms of $f(R)$ gravity. These corrected metric solutions include terms that reflect the modifications to classical black hole solutions due to $f(R)$ gravity. The thin shell model has been used to calculate the thermodynamic quantities for these black holes, showing the effects of different $f(R)$ forms on the horizon, entropy, surface gravity, and temperature.

This paper investigates the thermodynamic properties of Schwarzschild and Kerr-Newman black holes within various $f(R)$ gravity frameworks. Using the thin shell model, we derive expressions for entropy, temperature, and surface gravity for these modified black holes. Additionally, we explore the compatibility of these solutions with the scalar-Einstein solution proposed by Mark D. Roberts in 1985, providing a potential counterexample to the cosmic censorship hypothesis.


\begin{thebibliography}{99}
\bibitem[1]{1} J. D. Bekenstein, Phys. Rev. D 7 (1973) 2333 .

\bibitem[2]{2} S. W. Hawking, Nature 248 (1974) 30 .

\bibitem[3]{3} P. C. W. Davies, Proc. Roy. Soc. Lond. A $353,499(1977)$.

\bibitem[4]{4} R. G. Cai, L. M. Cao and Y. W. Sun, JHEP $11,039(2007)$.

\bibitem[5]{5} R. Penrose, Revista Del Nuovo Cimento, 1, 252 (1969).

\bibitem[6]{6}Griffin, Allan, David W. Snoke, and Sandro Stringari, eds. Bose-einstein condensation. Cambridge University Press, 1996.

\bibitem[7]{7}Chen, Wen-Xiang, and Yao-Guang Zheng. ``Thermodynamic geometric analysis of 3D charged black holes under f (R) gravity." arXiv preprint arXiv:2312.10043 (2023).

\bibitem[8]{8}Wei, Shao-Wen, and Yu-Xiao Liu. ``Insight into the microscopic structure of an AdS black hole from a thermodynamical phase transition." Physical review letters 115.11 (2015): 111302. 

\bibitem[9]{9}
Chen, Wen-Xiang. ``Thermodynamic Topology of Quantum RN Black Holes." arXiv preprint arXiv:2405.04541 (2024).
 
 \bibitem[10]{10}Roberts, Mark D. : Scalar Field Counter-Examples to the Cosmic Censorship Hypothesis. Gen.Rel.Grav.21(1989)907-939.

\bibitem[11]{11}Xiang L, Zheng Z. Entropy of a Vaidya black hole[J]. Physical Review D, 2000, 62(10): 104001.

\bibitem[12]{12} S. Soroushfar, R. Saffari and N. Kamvar, Eur. Phys. J. C 76,476 (2016).

\bibitem[13]{13} Taeyoon Moon, Yun Soo Myung, and Edwin J. Son. f(R) black holes. Gen. Rel. Grav., 43:3079-3098, $2011 .$

\bibitem[14]{14}Ahmad Sheykhi. Higher-dimensional charged $f(R)$ black holes. Phys. Rev., D86:024013, $2012 .$

\bibitem[15]{15}Engle, Jonathan, et al. ``The SU (2) black hole entropy revisited." Journal of High Energy Physics 2011.5 (2011): 1-30.

\bibitem[16]{16}Capozziello, Salvatore, et al. ``Curvature quintessence matched with observational data." International Journal of Modern Physics D 12.10 (2003): 1969-1982.

\bibitem[17]{17}Caravelli, Francesco, and Leonardo Modesto. ``Holographic effective actions from black holes." Physics Letters B 702.4 (2011): 307-311.

\bibitem[18]{18}Hendi, S. H., B. Eslam Panah, and S. M. Mousavi. ``Some exact solutions of F (R) gravity with charged (a) dS black hole interpretation." General Relativity and Gravitation 44 (2012): 835-853.

\bibitem[19]{19}Hu, Ya-Peng, Feng Pan, and Xin-Meng Wu. ``The effects of massive graviton on the equilibrium between the black hole and radiation gas in an isolated box." Physics Letters B 772 (2017): 553-558.

\bibitem[20]{20}T. Multamaki, I. Vilja, Spherically symmetric solutions of modified field equations in $f(R)$ theories of gravity[J]. Physical Review D, 2006, 74(6): 064022.
\end{thebibliography}
\end{document}